\documentclass{preprint}

\usepackage{tabularx}
\usepackage{courier}

\usepackage{xcolor}
\usepackage{booktabs}
\usepackage{graphicx} 
\usepackage[numbers,comma,sort&compress]{natbib}
\usepackage[colorlinks=true]{hyperref}

\title{Evaluating GPT-4V (GPT-4 with Vision) on Detection of Radiologic Findings on Chest Radiographs}
\author[1]{Yiliang Zhou, MS}
\author[2]{Hanley Ong, MD}
\author[2]{Patrick Kennedy, MD}
\author[3]{Carol Wu, MD}
\author[2]{Jacob Kazam, MD}
\author[2]{Keith Hentel, MD, MS}
\author[4]{Adam Flanders, MD}
\author[2,*]{George Shih, MD}
\author[1,*]{Yifan Peng, PhD}
\affil[1]{Department of Population Health Sciences, Weill Cornell Medicine, New York, NY 10065}
\affil[2]{Department of Radiology, Weill Cornell Medicine, New York, NY 10065}
\affil[3]{Department of Thoracic Image, University of Texas MD Anderson Cancer Center, Houston, TX 77030}
\affil[4]{Department of Radiology, Thomas Jefferson University Hospital, Philadelphia, PA 19107}
\affil[*]{Corresponding: \url{ges9006@med.cornell.edu}, \url{yip4002@med.cornell.edu}}

\begin{document}

\maketitle

\begin{abstract}
\textbf{Background}

Generating radiologic findings from chest radiographs is pivotal in medical image analysis. The emergence of OpenAI's generative pretrained transformer, GPT-4 with vision (GPT-4V)\cite{OpenAI2023GPT4V}, has opened new perspectives on the potential for automated image-text pair generation. However, the application of GPT-4V to real-world chest radiography is yet to be thoroughly examined.

\textbf{Purpose}

To investigate GPT-4V's capability to generate radiologic findings from real-world chest radiographs.

\textbf{Materials and Methods}

In this retrospective study, 100 chest radiographs with free-text radiology reports were annotated by a cohort of radiologists, two attending physicians and three residents, to establish a reference standard. Out of 100 chest radiographs, 50 were randomly picked from the National Institutes of Health (NIH) chest radiography data set, and 50 were randomly selected from the Medical Imaging and Data Resource Center (MIDRC). The performance of GPT-4V at detecting imaging findings from each chest radiograph was assessed in the zero-shot settings (where it operates without prior examples) and few-shot settings (where it operates with two examples). Its outcomes were compared with the reference standard about clinical conditions and their corresponding codes in the \textit{International Classification of Diseases, Tenth Revision} (ICD-10 codes), including the anatomic location (hereafter, laterality).

\textbf{Results}

In the zero-shot setting, in the task of detecting ICD-10 codes alone, GPT-4V attained an average PPV of 12.3\%, an average TPR of 5.8\%, and an average F1 score of 7.3\% on the NIH data set, and an average PPV of 25.0\%, an average TPR of 16.8\%, and an average F1 score of 18.2\% on the MIDRC data set. When both the ICD-10 codes and their corresponding laterality were considered, GPT-4V produced an average PPV of 7.8\%, an average TPR of 3.5\%, and an average F1 score of 4.5\% on the NIH data set, and an average PPV of 3.6\%, an average TPR of 4.9\%, and an average F1 of 6.4\% on the MIDRC data set. With few-shot learning, GPT-4V showed improved performance on both data sets. When contrasting zero-shot and few-shot learning, there were improved average TPRs and F1 scores in few-shot setting. Nonetheless, there was not a substantial increase in the average PPV.

\textbf{Conclusion}

Although GPT-4V has shown promise in understanding nature images, it had limited effectiveness in interpreting real-world chest radiographs.
\end{abstract}

\section{Introduction}\label{introduction}

Generating radiologic findings from chest radiographs is pivotal in medical image analysis\cite{Speets2006-mr}. Recent advancements in fine-tuned pretrained models have showcased their capability to translate image content into text\cite{Yang2023-sh}. However, these models are often trained on extensive nonspecific data sets and may need more domain-specific tuning for chest radiographs. The emergence of OpenAI's generative pretrained transformer, GPT-4 with vision (GPT-4V), a multimodal large language model (LLM) with visual recognition, has opened new perspectives on the potential for automated image-text pair generation in the medical care domain. Advanced multimodal LLMs, such as GPT-4V, can understand both text and images. While several studies have investigated the performance of GPT-4 in generating radiologic impressions\cite{Wang2017-ChestX} and summarizing clinical trials\cite{Sun2023-Evaluating}, the practical application of multimodal LLMs to real-world chest radiographs is yet to be thoroughly examined. Motivated by this knowledge gap, this study aimed to investigate the capability of GPT-4V to generate radiologic findings from real-world chest radiographs.

\section{Materials \& Methods}\label{materials-methods}

Because of the publicly available nature of the data set used in this study, the requirement to obtain written informed consent from all subjects was waived by the institutional review board.

\subsection{Study Design and Data Collection}

\begin{figure}
    \centering
    \includegraphics[width=.8\textwidth]{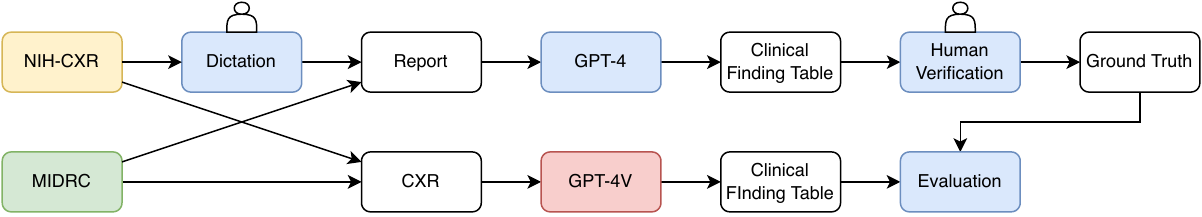}
    \caption{Diagram shows the study workflow, including construction of data and application of GPT-4 and GPT-4 with vision (GPT-4V). CXR= chest radiograph, MIDRC= Medical Imaging and Data Resource Center, NIH= National Institutes of Health}
    \label{fig:workflow}
\end{figure}

In this retrospective study, a total of 100 chest radiographs and radiology reports were independently annotated by a cohort of radiologists that included two attending physicians and three residents to establish a reference standard (Fig \ref{fig:workflow}). Of 100 chest radiographs, 50 were randomly selected from the National Institutes of Health (NIH) chest radiography data set, and their corresponding reports were dictated by one radiology attending physician and three radiology residents\cite{Wang2017-ChestX}. These 50 patients have been previously reported\cite{Sun2023-Evaluating}. The prior article dealt with the generation of the impression section by using the findings section in the report, whereas the current article describes the generation of a table of radiologic finding from the image.

The remaining 50 chest radiographs and de-identified free-text radiology reports were randomly selected from the Medical Imaging and Data Resource Center (MIDRC)\cite{Sun2023-Evaluating}. Each report included a findings section and an impressions section.

Out of 100 chest radiographs and reports, 10 cases were randomly selected (five from the NIH data set and five from the MIDRC data set), and two were randomly selected to serve as few-shot examples for the GPT-4V model. The remaining 90 were used to evaluate the performance of GPT-4V in zero-shot learning setting, where it operates without prior examples, and in few-shot learning setting, where it operates with two examples. Its outcomes were then compared with the reference standard annotated by radiologists with regard to clinical conditions, their corresponding codes in the \textit{International Classification of Diseases, Tenth Revision} (ICD-10), including laterality.

\subsection{GPT-4 with Vision (GPT-4V)}

GPT-4V (Oct.13th, 2023 version; OpenAI) was used in this study\cite{OpenAI2023GPT4V}. GPT-4V is a version of GPT-4 that allows users to instruct the LLM to analyze image inputs.

\subsection{Experimental Setup}\label{experimental-setup}

\begin{figure}
    \centering
    \includegraphics[width=\textwidth]{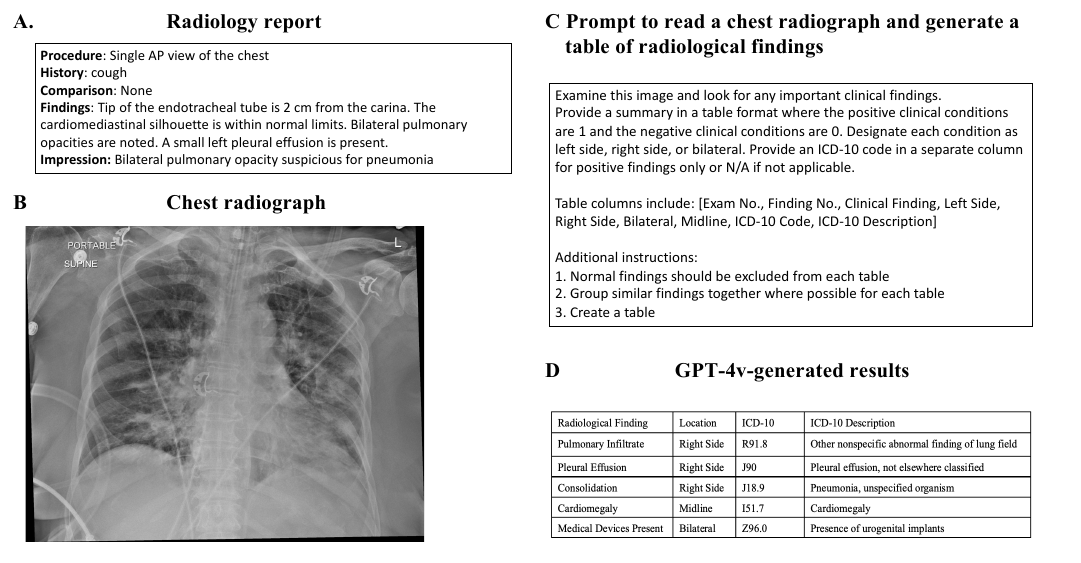}
    \caption{An example of GPT-4 with vision (GPT-4V) inputs and output, including the \textbf{(A)} radiology report, \textbf{(B)} chest radiograph, \textbf{(C)} prompt provided to GPT-4V to create a table of radiologic findings derived from the chest radiograph, and \textbf{(D)} resultant table of radiographic findings generated by GPT-4V. AP = anteroposterior.}
    \label{fig:example}
\end{figure}

To obtain the reference standard tables, GPT-4 was used to convert each free-text radiology report (Fig \ref{fig:example}A) into a table of radiologic findings by using a textual prompt (Appendix \ref{sec:prompt}). This table includes the radiologic findings, the corresponding ICD-10 diagnostic codes, and their laterality, as well as detailed descriptions of the ICD codes (Table~S\ref{tab:reference}). Subsequently, each report was independently evaluated by three readers from a cohort of five board-certified radiologists and residents (H.O., P.K., C.W., J.K., G.S). Two of the readers were 3rd-year radiology residents and the remaining three are radiology attending physicians, each with over 15 years of experience. Their radiology subspecialties cover chest, emergency department, bone, neurology, and body imaging. All readers had access to the image views and reports but not to additional clinical or patient data. Both data sets, the National Institutes of Health (NIH) and the Medical Imaging and Data Resource Center (MIDRC), were comprehensively reviewed, with the 50 reports of each data set being examined by three readers to maintain consistency and objectivity in the evaluation process. Findings were only included in the final tables if they were observed by at least two of the three readers; conversely, findings were excluded if two or more readers did not identify them. The majority vote principle was employed to provide a clear consensus for the presence or absence of radiologic findings and lead to the final reference standard table for each radiograph.

\subsubsection{Evaluating Performance in the Zero-Shot Setting}

To evaluate performance of GPT-4V in the zero-shot setting, the chest radiographs (Fig \ref{fig:example}B) were input into the GPT-4V model (accessed Oct.13\textsuperscript{th}, 2023)\cite{OpenAI2023GPT4V}\cite{Yang2023-sh} along with the prompt (Fig \ref{fig:example}C). The aim of this step was to generate a radiologic findings table (Fig \ref{fig:example}D) comparable to the reference standard. The analysis concentrated on aligning positive radiologic finding identification and laterality between GPT-generated tables and the consensus tables. The positive predictive value (PPV), true positive rate (TPR), and F1 score were calculated at the report level (see Statistical Analysis). Notably, the per-report average F1 scores used in this study as each case could have multiple diagnoses. These metrics were used to assess the accuracy of GPT-4V in detecting the ICD-10 codes and their respective laterality.

\subsubsection{Evaluating Performance in the Few-Shot Setting}

To evaluate the performance of GPT-4V in the few-shot setting, the input was extended to include two examples of chest radiographs with their corresponding radiologic findings tables before the prompt. From the pool of 10 chest radiographs, two were randomly selected to serve as few-shot examples for the GPT-4V model. Supplying the model with these examples helped to provide context, boosting the model's capacity to generate an accurate radiologic findings table. The same performance metrics (PPV, TPR, F1 score) were used at the report level to assess the effectiveness of few-shot learning.

\subsection{Statistical Analysis}\label{statistical-analysis}

When obtaining the final reference standard tables, the interrater agreement was assessed using Cohen's Kappa coefficient\cite{Waisberg2023-xy}.

First, the GPT-4V was employed to detect radiologic findings from each chest radiography and these results were compared with the predicted findings obtained from the radiologists. To convey the performance evaluation, PPV was used to denote the proportion of ICD-10 codes correctly predicted by GPT-4V, while the TPR represented the ratio of true-positive predictions to the total number of ICD-10 codes identified by GPT-4V.

PPV, TPR, and F1 score were the metrics used to assess the performance of GPT-4V in detecting imaging findings from each chest radiograph. The F1 socre is the harmonic mean of the PPV and TPR. These predicted findings were then compared with those obtained from the radiologists using the following equation: 

\begin{equation}
    F1 = 2 \times \frac{PPV \times TPR}{PPV + TPR}
\end{equation}

In the evaluation of detecting ICD-10 codes, a radiologic finding was considered a true-positive if its ICD-10 code aligned with that in the reference standard table. In evaluating both the radiologic findings in ICD-10 codes and their corresponding lateralities, a radiologic finding was considered a true positive if both its ICD-10 code and laterality matched those in the reference standard table.

For example, when evaluating ICD-10 codes alone, there were two ICD-10 Codes correctly predicted by GPT-4V (J90 and J18.9), with five findings predicted by GPT-4V (Fig \ref{fig:example}), and four in the reference standard table (Table S1). Therefore, PPV was 0.4 ($=\frac{2}{2+3}$) , TPR was 0.5 ($=\frac{2}{2+2}$), and F1 score was 0.44.

After obtaining the PPV, TPR, and F1 score for each chest radiography, the macro averages were calculated for PPV, TPR, and F1 score across all 90 chest radiographs.

$P = .05$ was considered indicative of a statistically significant difference. Two-tailed t-tests were used for calculating the \textit{P} values for the performance metrics of GPT-4V in detecting radiologic findings from chest radiographs, specifically assessing the detection of findings represented by their associated lateralities and ICD-10 codes in both zero-shot and few-shot settings (Figs \ref{fig:zero}-\ref{fig:diff}).

\section{Results}\label{results}

\subsection{Performance in the Zero-Shot Setting}

\begin{figure}[tbph]
    \centering
    \includegraphics[width=\textwidth]{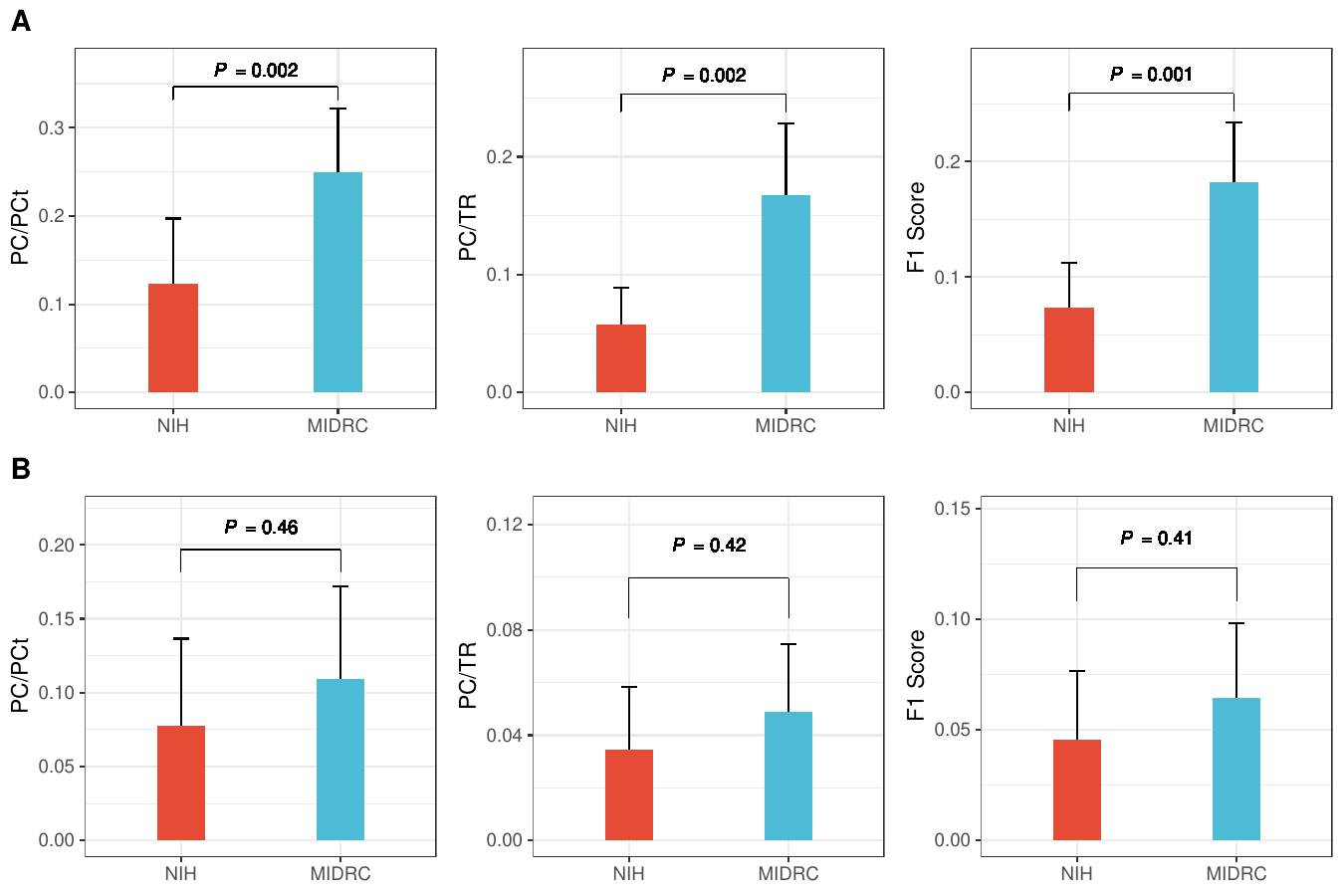}
    \caption{Bar graphs show the performance of GPT-4 with vision (GPT-4V) in the detection of radiologic findings from chest radiographs in the zero-shot setting, with statistical significance assessed using the two-tailed t-test, according to \textbf{(A)} the radiologic findings in \textit{International Statistical Classification of Diseases, Tenth Revision} (ICD-10) codes only and \textbf{(B)}  both the radiologic findings in ICD-10 codes and their corresponding lateralities.}
    \label{fig:zero}
\end{figure}

The performance of the GPT-4V model in the zero-shot setting varied across the NIH and MIDRC data sets and for different circumstances (Fig \ref{fig:zero}). In the task of detecting ICD-10 codes alone, the model attained an average PPV of 5.53 of 45 remaining radiographs(12.3\%) (SD [0.25], SE [0.04], IQR [0.20]), average TPR of 2.60 of 45 (5.8\%) (SD [0.10], SE [0.02], IQR [0.10]), and average F1 score of 3.30 of 45 (7.3\%) (SD [0.13], SE [0.02], IQR [0.14]) on the NIH data set. Conversely, on the MIDRC data set, the model managed an average PPV of 11.25 of 45 (25.0\%) (SD [0.24], SE [0.04], IQR [0.33]), average TPR of 7.56 of 45 (16.8\%) (SD [0.20], SE [0.03], IQR [0.25], and average F1 score of 8.20 of 45 (18.2\%) (SD [0.17], SE [0.03], IQR [0.29]). The notable differences in measurements between the two data sets were primarily due to the MIDRC data set having fewer missing GPT-4V–generated ICD-10 codes than the NIH data set. On the MIDRC data set, GPT-4V generated 144 radiologic findings, while the reference standard comprised 261 findings. However, on the NIH data set, GPT-4V produced 102 radiologic findings, while the reference standard comprised 220 findings. Nevertheless, when both the ICD-10 codes and their corresponding laterality were taken into account, the GPT-4V model in the zero-shot setting produced an average PPV of 3.5 of 45 (7.8\%) (SD [0.20], SE [0.03], IQR [0.0], average TPR of 1.56 of 45 (3.5\%) (SD [0.01], SE [0.01], IQR [0.0]), and F1 score of 2.05 of 45 (4.5\%) (SD [0.10], SE [0.02], IQR [0.0]) on the NIH data set; and an average PPV of 4.92 of 45 (10.9\%) (SD [0.21], SE [0.03], IQR [0.25]), average TPR of 2.19 of 45 (4.9\%) (SD [0.09], SE [0.01], IQR [0.10]), and average F1 of 2.90 of 45 (6.4\%) (SD [0.11], SE [0.02], IQR [0.15]) on the MIDRC data set.

\subsection{Performance in the Few-Shot Setting}

\begin{figure}[tbph]
    \centering
    \includegraphics[width=\textwidth]{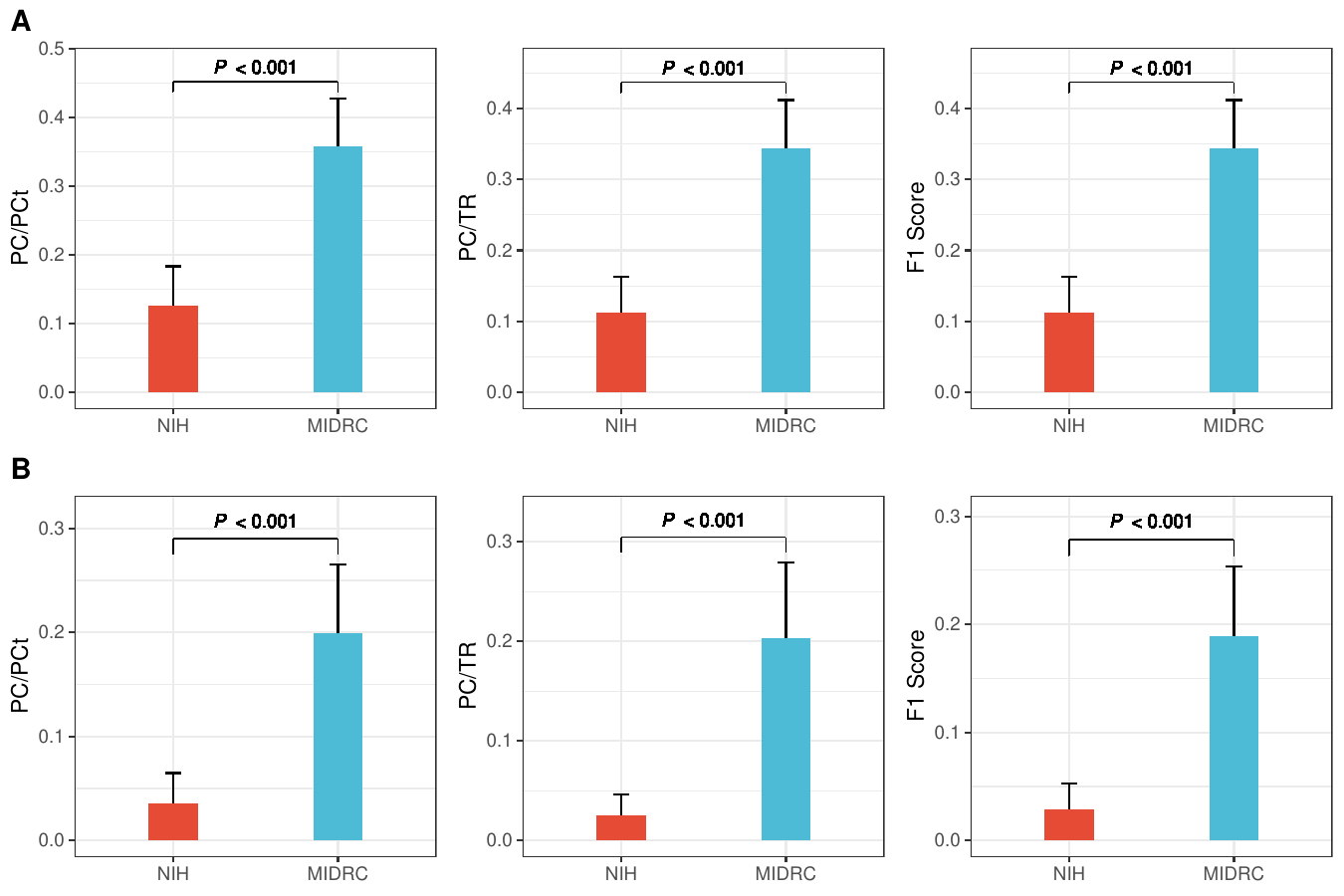}
    \caption{Bar graphs show the performance of GPT-4 with vision (GPT-4V) in the detection of radiologic findings from chest radiographs in the few-shot setting, with statistical significance assessed using the two-tailed t test, according to \textbf{(A)} the radiologic findings in \textit{International Statistical Classification of Diseases, Tenth Revision} (ICD-10) codes only and \textbf{(B)}  both the radiologic findings in ICD-10 codes and their corresponding lateralities.}
    \label{fig:few}
\end{figure}

With few-shot learning, GPT-4V showed improved performance on both NIH and MIDRC data sets (Fig~\ref{fig:few}). When the model was provided with two illustrative chest radiographs and their corresponding radiologic finding tables, there was a marked improvement in the average PPV on the NIH data set, with rates rising to 5.72/45 remaining radiographs (12.7\%) (SD [0.19], SE [0.03], IQR [0.25]). The average TPR also enhanced to 4.69 of 45 (10.4\%) (SD [0.17], SE [0.03], IQR [0.22]), while the average F1 score reached 5.03 of 45 (11.1\%) (SD [0.17], SE [0.03], IQR [0.22]). On the MIDRC data set, the average PPV improved to 16.15 of 45 (35.9\%) (SD [0.23], SE [0.03], IQR [0.50]), the average TPR improved to 16.68 of 45 (37.1\%) (SD [0.23], SE [0.03], IQR [0.50]), and the average F1 score improved to 15.47 of 45 (34.3\%) (SD [0.23], SE [0.03], IQR [0.50]). When tasked with detecting both ICD-10 codes and their corresponding lateralities, the GPT-4V model demonstrated improved efficacy. On the NIH data set, it displayed an average PPV of 1.62 of 45 (3.5\%) (SD [0.10], SE [0.01], IQR [0.0]), average TPR of 1.14 of 45 (2.5\%) (SD [0.07], SE [0.01], IQR [0.0]), and average F1 score of 1.30 of 45 (2.8\%) (SD [0.08], SE [0.01], IQR [0.0]). On the MIDRC data set, GPT-4V achieved an average PPV of 8.96 of 45 (19.9\%) (SD [0.22], SE [0.03], IQR [0.33]), average TPR of 9.14 of 45 (20.3\%) (SD [0.25], SE [0.04], IQR [0.33]), and average F1 score of 8.53 of 45 (19.0\%) (SD [0.21], SE [0.03], IQR [0.31]).

\begin{figure}
    \centering
    \includegraphics[width=\textwidth]{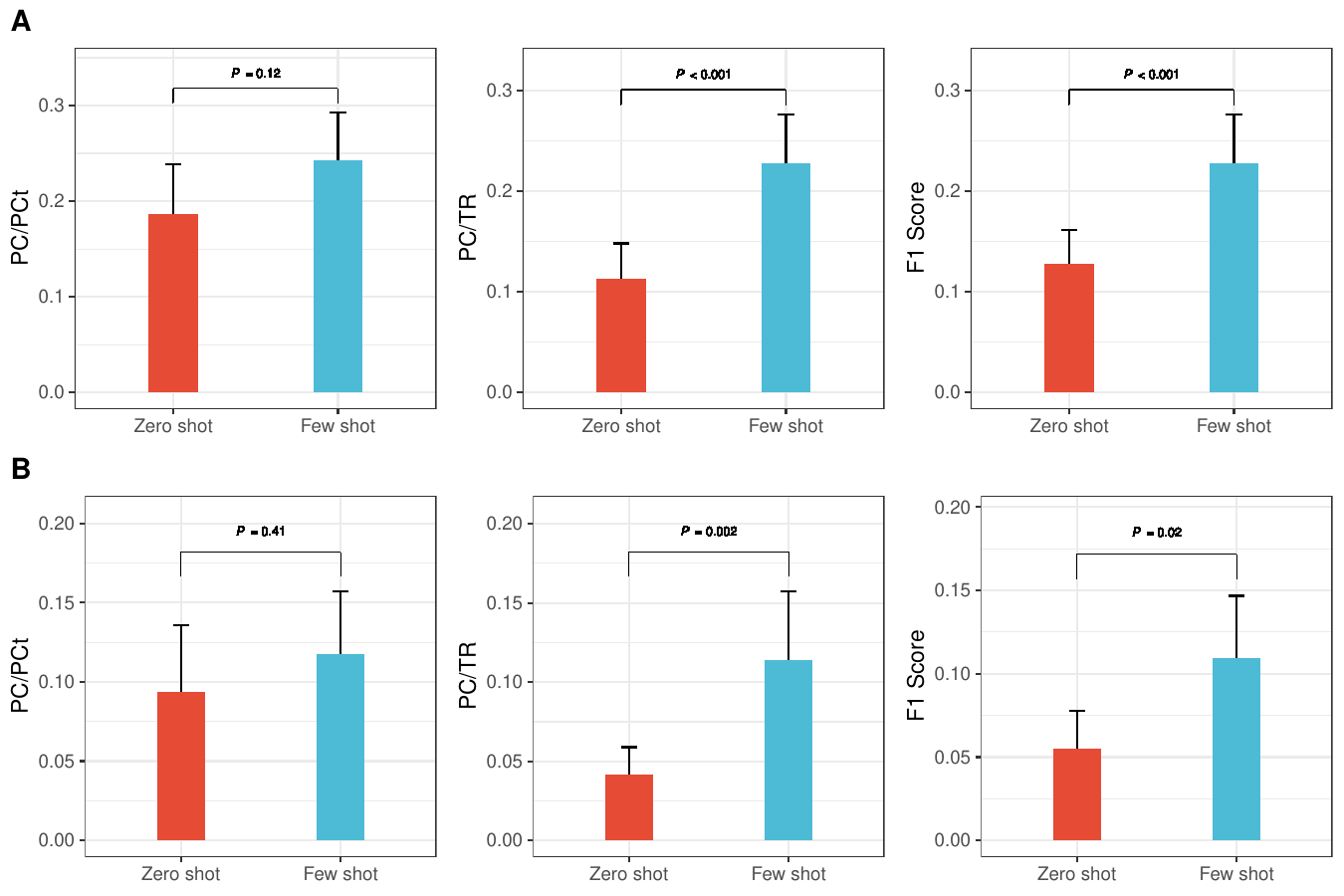}
    \caption{Bar graphs show the difference in performance of GPT-4 with vision (GPT-4V) in the detection of radiologic findings from chest radiographs between the zero-shot and few-shot settings, with statistical significance assessed using the two-tailed t-test, according to \textbf{(A)} the radiologic findings in \textit{International Statistical Classification of Diseases, Tenth Revision} (ICD-10) codes only and \textbf{(B)}  both the radiologic findings in ICD-10 codes and their corresponding lateralities.}
    \label{fig:diff}
\end{figure}

When contrasting zero-shot and few-shot learning approaches on both data sets (Fig \ref{fig:diff}), there were improved average TPR and F1 scores with few-shot learning in both scenarios (ICD-10 codes only and ICD-10 codes with laterality). Nonetheless, there was not a substantial increase in the average PPV, suggesting that while few-shot learning may enhance the model's capacity to detect findings, it does not noticeably enhance the precision of the predictions.

\subsection{Inter-rater Agreement}

The study compared each reader's outcomes to the reference standard to evaluate their interpretations (Table~\ref{tab:ira}). On the NIH data set, the interrater agreement was 0.78 for reader A, -0.06 for reader C, 0.96 for reader D. On the MIDRC data set, the interrater agreement was 0.96 for reader B, 0.82 for reader C, and 0.99 for reader D.

\begin{table}[hbpt]
\vspace{1em}
\caption{Reader Agreement for NIH and MIDRC.
Note. ---Interrater Agreement between each reader and the reference standard table for both NIH and MIDRC data sets were accessed with Cohen $\kappa$ statistics. $\kappa$ values were interpreted as follows: \(\kappa = 1\) indicates perfect agreement among reviewers and the reference standard; \(\kappa = 0\) indicates that agreement is no better than chance; \(\kappa = \  - 1\) indicates perfect disagreement.MIDRC = Medical Imaging and Data Resource Center, NA = not applicable, NIH = National Institutes of Health.}
\label{tab:ira}
\centering
\begin{tabular}{lrr}
\toprule
& NIH $\kappa$ Score & MIDRC $\kappa$  Score \\
\midrule
Reader A & 0.78 & NA \\
Reader B & NA & 0.96 \\
Reader C & -0.06 & 0.82 \\
Reader D & 0.96 & 0.99 \\
\bottomrule
\end{tabular}
\end{table}

\section{Discussion}\label{discussion}

The emergence of multimodal large language models (LLMs) that can understand both text and images, such as OpenAI’s GPT-4V \cite{OpenAI2023GPT4V}, shows potential for automated image-text pair generation. However, applying these models to real-world data is yet to be thoroughly examined. This study assessed the feasibility of using GPT-4V to detect radiologic findings from chest radiographs in both zero-shot and few-shot learning contexts. The results (average PPV, 5.53 of 45 [12.3\%]; average TPR, 2.60 of 45 [5.8\%]) demonstrated that radiologic findings tables generated by GPT-4V still need further preparation for use in clinical practice. We acknowledge a limitation in employing GPT-4 for converting radiologic reports into a structured table, where inaccuracies in the International Statistical Classification of Diseases, Tenth Revision (ICD-10) code assignments and distinguishing between radiologic findings and conclusions may impact the reliability and interpretability of the data. A notable limitation of the GPT-4V output was its failure to detect several clinical conditions based on corresponding ICD-10 codes. Overall, the top three findings that GPT-4V could not detect were “endotracheal tube,” “central venous catheter,” and “degenerative changes of osseous structures.” Conversely, GPT-4V most accurately detected findings such as “chest drain,” “air-space disease,” and “lung opacity.”

Although GPT-4V showed promise in understanding real-world images\cite{Zhang2023-ax}, its effectiveness in interpreting real-world chest radiographs was limited.

Our study had limitations. First, few-shot training is that it may be more prone to generating ICD-10 codes already in the provided examples, potentially reducing the diversity of ICD-10 codes generated. Second, we were unable to access other multimodal LLMs that support image inputs. Consequently, we lacked comparative data with respect to the results of GPT-4v. Finally, due to the limited size of the data set and relatively low level of inter-rater agreement among the radiologists, analysis by GPT-4V may have been challenging.

In conclusion, GPT-4V has shown promise in understanding natural images but had limited effectiveness in interpreting real-world chest radiographs. Our results highlight the need for additional comprehensive development and assessments prior to incorporating the GPT-4V model into clinical practice routines. Task-specific, fine-tuned multimodal LLMs or foundation models are urgently needed for this purpose, although it is not necessarily the best solution. To yield robust and generalizable results, we plan to explore larger and more diverse data sets using real-world data in future studies. This will involve including multiple modalities, such as brain CT scans and MRIs, to conduct a more thorough evaluation of the performance of GPT-4V.

\section*{Acknowledgment}

This work was supported by the National Institutes of Health under Award No. 4R00LM013001 (Peng), NSF CAREER Award No. 2145640 (Peng), and Amazon Research Award (Peng, Shih).

We acknowledge parts of this article were generated with GPT-4V (powered by OpenAI’s language model; \url{https://openai.com}).

\bibliographystyle{medline}
\bibliography{ref}

\begin{thebibliography}{1}

\bibitem{OpenAI2023GPT4V}
{OpenAI}. GPT-4V.
\newblock \url{https://openai.com/research/gpt-4v-system-card}. Accessed October 13, 2023.

\bibitem{Speets2006-mr}
Speets AM, van~der Graaf Y, Hoes AW, Kalmijn S, Sachs AP, Rutten MJ, Gratama JWC, Montauban~van Swijndregt AD, Mali WP.
\newblock Chest radiography in general practice: indications, diagnostic yield and consequences for patient management.
\newblock Br J Gen Pract. 2006 Aug;56(529):574--578.
\newblock PMID: 16882374. PMCID: PMC1874520.

\bibitem{Yang2023-sh}
Yang Z, Li L, Lin K, Wang J, Lin CC, Liu Z, Wang L.
\newblock The Dawn of {LMMs}: Preliminary Explorations with {GPT-4V(ision}).
\newblock ArXiv. 2023 Sep;abs/2309.17421.
\newblock doi: 10.48550/arXiv.2309.17421. Accessed on: November 15, 2023.

\bibitem{Wang2017-ChestX}
Wang X, Peng Y, Lu L, Lu Z, Bagheri M, Summers RM.
\newblock {ChestX-Ray8}: {Hospital-Scale} Chest {X-Ray} Database and Benchmarks on {Weakly-Supervised} Classification and Localization of Common Thorax Diseases.
\newblock In: 2017 {IEEE} Conference on Computer Vision and Pattern Recognition ({CVPR}); 2017. p. 3462--3471.
\newblock doi: 10.1109/CVPR.2017.369.

\bibitem{Sun2023-Evaluating}
Sun Z, Ong H, Kennedy P, Tang L, Chen S, Elias J, Lucas E, Shih G, Peng Y.
\newblock Evaluating {GPT4} on Impressions Generation in Radiology Reports.
\newblock Radiology. 2023 Jun;307(5):e231259.
\newblock doi: 10.1148/radiol.231259. PMID: 37367439.

\bibitem{Waisberg2023-xy}
Waisberg E, Ong J, Masalkhi M, Kamran SA, Zaman N, Sarker P, Lee AG, Tavakkoli A.
\newblock {GPT-4}: a new era of artificial intelligence in medicine.
\newblock Ir J Med Sci. 2023 Dec;192(6):3197--3200.
\newblock doi: 10.1007/s11845-023-03377-8. PMID: 37076707.

\bibitem{Zhang2023-ax}
Zhang X, Lu Y, Wang W, Yan A, Yan J, Qin L, Wang H, Yan X, Wang WY, Petzold LR.
\newblock {GPT-4V(ision}) as a Generalist Evaluator for {Vision-Language} Tasks.
\newblock ArXiv. 2023 Nov.
\newblock Accessed on: November 15, 2023.

\end{thebibliography}

\newpage
\appendix
\setcounter{table}{0}
\setcounter{figure}{0}
\renewcommand\figurename{Supplementary Figure} 
\renewcommand\tablename{Supplementary Table}


\section{Prompt to convert the free-text report into a table of radiological findings}
\label{sec:prompt}

{\ttfamily Read the report and look for any important clinical findings. Provide a summary in a table format where the positive clinical conditions are 1 and the negative clinical conditions are 0. Designate each condition as left side, right side, or bilateral, Provide an ICD-10 code in a separate column for positive findings only or N/A if not applicable.

Table columns include: [Exam No., Finding No., Clinical Finding, Left Side. Right Side, Bilateral, Midline, lCD-10 Code, iCD-10 Description]

Additional instructions:\\
1. Normal findings should be excluded from each table\\
2. Group similar findings together where possible for each table\\
3. Create a table\\
}

\newpage

\section{Reference standard radiological findings}

\begin{table}[!htbp]
\caption{Reference standard radiological findings}
\label{tab:reference}
\begin{tabularx}{\textwidth}{lllX}
\toprule
Radiologic Finding & Location & ICD-10 & ICD-10 Description \\
\midrule
Endotracheal tube placement & Midline & Z93.0 & Tracheostomy status \\
Pulmonary opacities & Bilateral & J84.9 & Interstitial pulmonary disease, unspecified\\
Pleural effusion C & Left Side & J90 & Pleural effusion, not elsewhere classified\\
Pneumonia & Bilateral & J18.9 & Pneumonia, unspecified organism \\
\bottomrule
\end{tabularx}
\end{table}

\end{document}